\documentclass[prd,twocolumn,floatfix,amsmath,nofootinbib,amssymb,floatfix]{revtex4}
\usepackage{graphicx,color,dcolumn,booktabs,bm}
\usepackage{longtable,lscape}
\usepackage{txfonts}
\usepackage{overpic}
\usepackage{amssymb}
\usepackage{indentfirst}
\usepackage{feynmf}   %{feynmp}
\usepackage{slashed}  %for Feynman symbols
\usepackage{cases}
\usepackage{color}
\usepackage{multirow}
\usepackage{epstopdf}
\usepackage{enumerate}
\usepackage{graphicx,color,dcolumn,booktabs,bm}
\usepackage[colorlinks, citecolor=blue,anchorcolor=red,menucolor=red, linkcolor=red,filecolor=red,urlcolor=blue,frenchlinks=red]{hyperref}

\begin{document}
%\begin{CJK}{GBK}{}

\title{Mapping a new cluster of charmoniumlike structures at $e^+e^-$ collisions}
\author{Jun-Zhang Wang$^{1,2}$}\email{wangjzh2012@lzu.edu.cn}
\author{Dian-Yong Chen$^3$}\email{chendy@seu.edu.cn}
\author{Xiang Liu$^{1,2,4}$\footnote{Corresponding author}}\email{xiangliu@lzu.edu.cn}
\author{Takayuki Matsuki$^{5}$}\email{matsuki@tokyo-kasei.ac.jp}
\affiliation{$^1$School of Physical Science and Technology, Lanzhou University, Lanzhou 730000, China\\
$^2$Research Center for Hadron and CSR Physics, Lanzhou University $\&$ Institute of Modern Physics of CAS, Lanzhou 730000, China\\
$^3$School of Physics, Southeast University, Nanjing 210094, China\\
$^4$Lanzhou Center for Theoretical Physics, Key Laboratory of Theoretical Physics of Gansu Province, and Frontiers Science Center for Rare Isotopes, Lanzhou University, Lanzhou 730000, China\\
$^5$Tokyo Kasei University, 1-18-1 Kaga, Itabashi, Tokyo 173-8602, Japan}

\date{\today}

\begin{abstract}
In this work, we find a {\it Critical Energy induced Enhancement} (CEE) mechanism for the general three-body open-charm process at the $e^+e^-$ collisions, which utilizes the peculiar kinematic behavior of the $e^+e^-$ annihilation process involving three-body final states. We present a general analysis of a three-body process $e^+e^-\to BC\to B(C\to DE)$. When the center-of-mass (CM) energy %point 
of the $e^+e^-$ collision satisfies a critical relation $\sqrt{s}=m_B+m_C$, there clearly exists the reflection peak of an intermediate $C$ state near the threshold of the invariant mass distribution of $m_{BE}$ or $m_{BD}$, whose formation is very sensitive to the CM energy. The reflection enhancement phenomenon induced at the critical energy means that a new cluster of charmoniumlike structures can be experimentally mapped.  Taking an example of $e^+e^-\to D_s^{*-}D_{s2}^*(2573)^+ \to  D_s^{*-}(D^0K^+)$, we further illustrate this novel phenomenon when $\sqrt{s}=4.680$ GeV. 
What is more important is that a series of optimal CM energy points to search for new charmoniumlike structures in three-body open-charm processes from $e^+e^-$ annihilation are suggested, which can be accessible at BESIII and further BelleII as a new research topic.

\end{abstract}

\maketitle

\noindent\textit{Introduction.}---Since 2003, dozens of charmoniumlike $XYZ$ states have been observed 
with the accumulation of experimental data with high precision, which form a new family in hadron spectroscopy. Study of charmoniumlike states is constantly improving our knowledge of  exotic hadronic matter and is deepening our understanding of non-perturbative behavior of strong interaction (see review articles \cite{Chen:2016qju,Liu:2019zoy,Guo:2017jvc,Olsen:2017bmm,Brambilla:2019esw} for getting recent progress). Until now, the exploration of charmoniumlike $XYZ$ state has become a frontier in science of matter. 

At present, these reported charmoniumlike states can be grouped into five clusters which involve several typical production processes like $B$ decay, $e^+e^-$ collision, and $\gamma\gamma$ fusion \cite{Chen:2016qju,Liu:2019zoy}.
Among them, $e^+e^-$ collision is an ideal platform to capture charmoniumlike states. For example, abundant $Y$ states \cite{Aubert:2005rm,Yuan:2007sj,Wang:2007ea,Aubert:2006ge,Pakhlova:2008vn,BESIII:2016adj,Ablikim:2016qzw} which result in "{\it Y problem}" \cite{Ablikim:2019hff} can be directly produced by $e^+e^-$ collisions, while two $X$ states \cite{Abe:2007jna,Abe:2007sya} are from double charmonium production in $e^+e^-$ collision. Here, the Belle, BaBar and, running BESIII experiments based on $e^+e^-$ collision are the main force of finding out charmoniumlike states. 
Obviously, $e^+e^-$ collision as an effective way to produce charmoniumlike states has played a crucial role when constructing charmoniumlike state families in the past 18 years.
Standing at a new starting point of new ten years, we believe that $e^+e^-$ collision will still be a rich mine of producing charmoniumlike states.

In this letter, we find a {\it Critical Energy induced Enhancement} (CEE) mechanism for the three-body open-charm processes from $e^+e^-$ annihilation, by which we map a new cluster of charmoniumlike structures that can be largely produced on the invariant mass distribution of the corresponding open-charm channel by choosing a specific critical energy of positron-electron system.  Different from possible hidden-charm exotic hadronic resonances, these new charmoniumlike structures with a non-resonant nature can be more easily identified due to  the existence of the unique CEE mechanism. In the forthcoming years, BESIII and Belle II will have great ability and opportunity to hunt for the predicted new cluster of charmoniumlike structures, for which the concrete suggestion of searching for them will be given. 

\noindent\textit{CEE mechanism.}---In 2014, the BESIII collaboration reported two charged charmoniumlike structures $Z_c(3885)^{\pm}$ and $Z_c(4025)^{\pm}$ in the three-body open-charm processes $e^+e^- \to \pi^{\pm}(D\bar{D}^*)^{\mp}$ \cite{Ablikim:2013xfr} and $e^+e^- \to \pi^{\pm}(D^*\bar{D}^*)^{\mp}$ \cite{Ablikim:2013emm}, respectively, whose peak positions are very close to the production thresholds of respective open-charm channels $D^{(*)}\bar{D}^*$. The two charged $Z_c$ structures are generally treated to be good candidates of exotic hidden-charm tetraquark states due to the neutral property of a charmonium meson \cite{Chen:2016qju,Liu:2019zoy}.  In Ref. \cite{Wang:2020axi}, the Lanzhou group proposed a totally different opinion to understand the structures of $Z_c(3885)^{\pm}$ and $Z_c(4025)^{\pm}$, both of which can be uniformly decoded as the reflection structures of the $P$-wave charmed meson $D_1(2420)$ involved in $e^+e^- \to D\bar{D}_1(2420) \to D\bar{D}^*\pi$ and $e^+e^- \to D^{*}\bar{D}_1(2420) \to D^{*}\bar{D}^*\pi$, respectively.  Later, this unique reflection explanation to $Z_c(3885)^{\pm}$ was further supported by the BESIII's measurement on the degree of asymmetry in the angular distribution of final states \cite{Wang:2020axi}. Building on previous studies, in this work, we find a CEE mechanism existing in the general open-charm process with three-body final states, which can widely induce this specific reflection peak phenomenon near a threshold.   Based on this mechanism, in addition to $Z_c(3885)^{\pm}$ and $Z_c(4025)^{\pm}$, we can expect that more charmoniumlike $Z_c$ structures  can be produced from the $e^+e^-$ annihilation process by setting their individual critical energy. And then a new cluster of charmoniumlike structures can be established.  In the following, we will illustrate it in detail.

%Here, the intermediate charmed meson $D_1(2420)$ is an off-shell state at the measured CM energy of $\sqrt{s}=4.26$ GeV by BESIII, which can just produce an obvious reflection peak near the threshold of $D^{(*)}\bar{D}^*$.

\begin{figure}[htbp]
	\includegraphics[width=8cm,keepaspectratio]{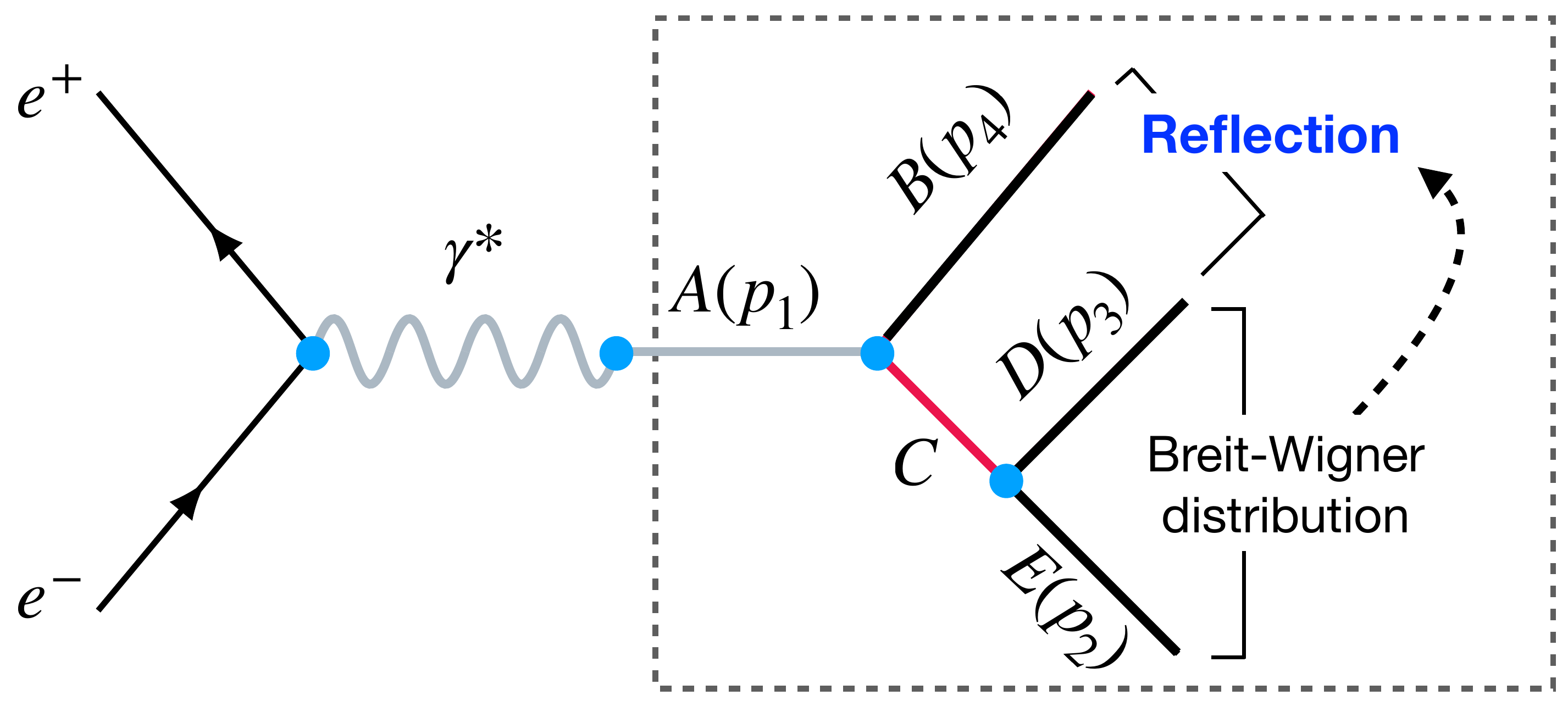}
	\caption{ The schematic diagram for a general three-body process from the electron and positron annihilation. Here, $A$ may denote a vector charmonium or charmoniumlike state with $J^{PC}=1^{--}$ coupling to a virtual photon.  $B$, $D$, and $E$ denote the particles involved in the final states, while $C$ is an intermediate state. $p_1$, $p_2$, $p_3$, and $p_4$ are defined to be the four-momenta of initial state and three final states, respectively, while $p_C=p_1-p_4$ corresponds to the propagation momentum of an intermediate $C$.  \label{fig:1}  }
\end{figure}

Focusing on the three-body open-charm channels from the $e^+e^-$ annihilation, $e^+e^- \to D^{(*)}\bar D_J^{**}\to D^{(*)}(\bar D^{(*)}\pi)$, $e^+e^- \to D^{(*)}\bar D_J^{**}\to  D^{(*)}(\bar D^{(*)}_sK)$, $e^+e^- \to D_s^{(*)}\bar D_{sJ}^{**}\to D_s^{(*)}(\bar D_s^{(*)}\eta)$, and  $e^+e^- \to D_s^{(*)}\bar D_{sJ}^{**}\to  D_s^{(*)}(\bar D^{(*)}K)$, we start with a general three-body process from the electron and positron annihilation, $A(p_1)\to BC\to B(p_4)(C\to D(p_3)E(p_2))$, whose schematic diagram can be seen in Fig.~\ref{fig:1}. Here, $D_{(s)J}^{**}$ denotes the higher states in the charmed/charmed-strange meson family. 
When choosing some special center-of-mass (CM) energy of the $e^+e^-$ collision ({$\sqrt{s}=m_B+m_C$}), the reflection signal of an intermediate $C$ clearly exists in the invariant mass distribution of $m_{BE}$ or $m_{BD}$. Especially, this reflection is sensitive to the width of  $C$ if $C$ is a narrow state.

\begin{figure}[h]
	\includegraphics[width=8.5cm,keepaspectratio]{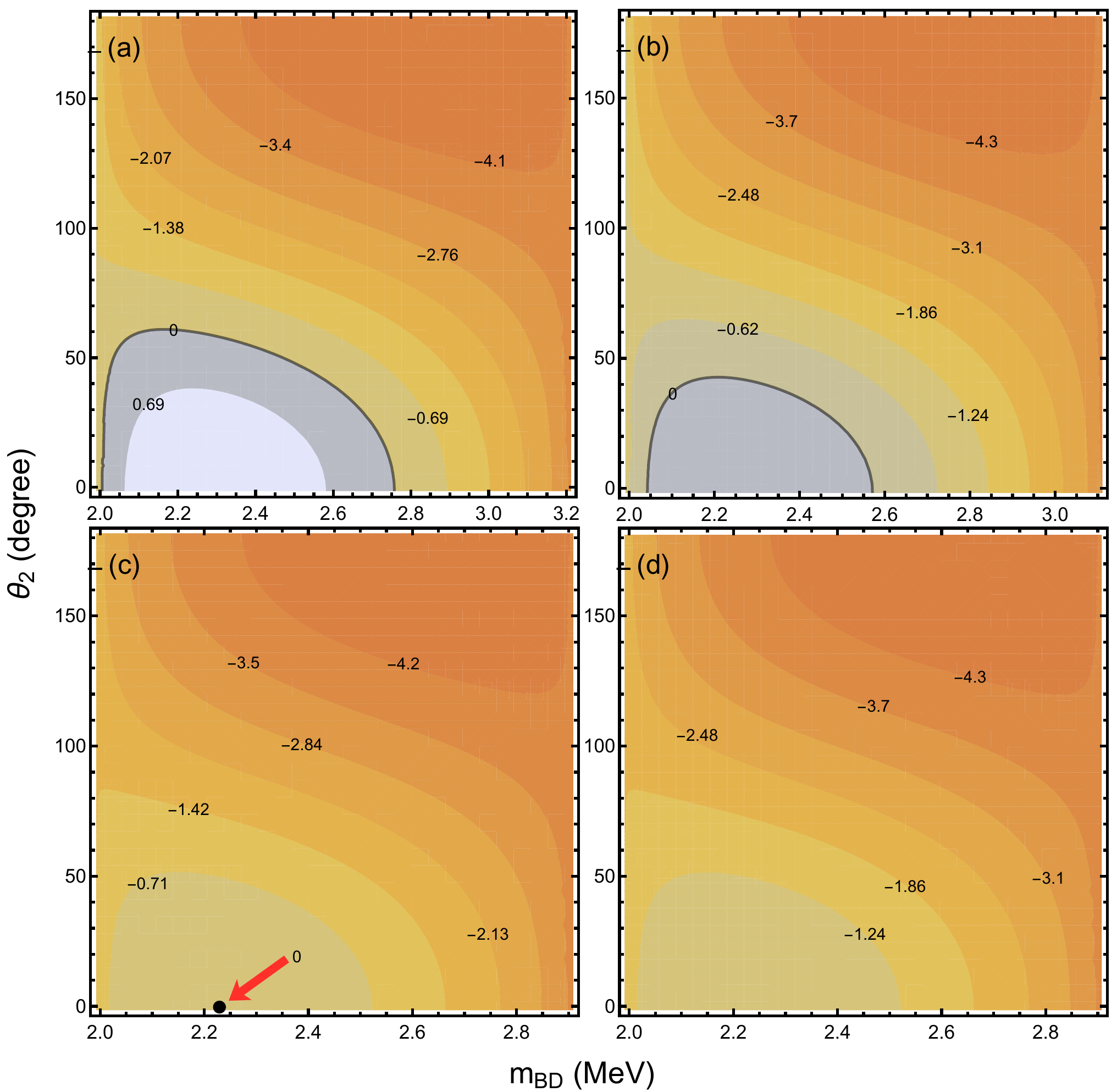}
	\caption{The contour plot of $(p_1-p_4)^2-m_C^2$ in the invariant mass $m_{BD}$ and scattering angle $\theta_2$. Here, we assume the masses of all final states to be 1 MeV and $m_C=3$ MeV. Diagrams ($a$)-($d$) correspond to $\sqrt{s}-\sqrt{s_{\textrm{critical}}}=200, 100, 0,  -100$  keV, respectively. 
 \label{fig:2}  }
\end{figure}

As we all know, in the three-body decay process shown in Fig. \ref{fig:1}, the pole information of an intermediate $C$ can be reflected on the indirect invariant mass spectrum of $m_{BE}$ or $m_{BD}$. However,  a kinematical behavior of the pole in indirect invariant mass spectrum is very different from that of direct invariant mass spectrum. Here, we take $m_{BD}$ as an example.  The pole information of an intermediate resonance is fully determined by the propagator $1/((p_1-p_4)^2-m_C^2+i m_C \Gamma_C)$ and is not dependent on the Lorentz structure of an interaction vertex. By solving the equation $(p_1-p_4)^2-m_C^2=0$, we can find a critical value $\sqrt{s_{\textrm{critical}}}=m_B+m_C$, which is a critical energy point of pole distributions of an intermediate state.  In Fig. \ref{fig:2} , we show the numerical contour plot of $(p_1-p_4)^2-m_C^2$ dependent on the invariant mass $m_{BD}$ and scattering angle $\theta_2$, where $\theta_2$ is the angle of three momentum of a particle $E$ to the $z$ axis in spherical coordinates. 
We notice that there appears a continuous pole distribution in the three-body phase space when the CM energy is larger than the critical energy.  When decreasing the CM energy close to the critical energy, the pole curve gradually tends to a single pole point, which exactly corresponds to the critical energy $\sqrt{s_{\textrm{critical}}}$. Of course, it is not surprising that there is no pole structure in the CM energy region smaller than $\sqrt{s_{\textrm{critical}}}$, 
%compared with the critical energy, 
where the intermediate state is always a virtual state. 

Therefore, this peculiar kinematic property near the critical energy point can cause some fantastic phenomenon, i.e., the line shape of a reflection peak for an intermediate $C$ is sensitive to the initial CM energy. In other word, a reflection enhancement structure near threshold that corresponds to a single pole can be largely induced at the CM energy point of $\sqrt{s}=\sqrt{s_{\textrm{critical}}}$, which is main meaning of the  CEE mechanism. The position of this single pole can be determined analytically with the following expression 
\begin{eqnarray}
m_{BD}^{pole}=\sqrt{\frac{(m_C^2-m_E^2+m_D^2)m_B+(m_D^2+m_B^2)m_C}{m_C}}, \label{eq:1}
\end{eqnarray} 
which is useful for the concrete experimental analysis. By inputting the mass of intermediate $C$, this pole position of a reflection peak structure can  be obtained. 

%BESIII start take data above 4.60 GeV

\textit{Typical example.}--To intuitively illustrate this novel phenomenon, we choose $e^+e^-\to D_s^{*-}D_{s2}^*(2573)^+ \to  D_s^{*-}(D^0K^+)$ as an example, where the intermediate $D_{s2}^*(2573)^+$ has $J^P=2^+$ spin-parity quantum number and is a narrow state \cite{Tanabashi:2018oca}. Since the sum of $m(D_{s2}^*(2573))=2.568$ GeV \cite{Tanabashi:2018oca,Aaij:2014xza} and $m(D_{s}^*)=2.112$ GeV \cite{Tanabashi:2018oca} is precisely equal to 4.680 GeV, we may set the CM energy of this discussed process to be $\sqrt{s}=4.68$ GeV. Considering the existent charmoniumlike state $Y(4660)$ \cite{Wang:2007ea,Wang:2020prx}, we may expect that $D_s^{*-}D_{s2}^*(2573)^+$ may couple with 
this $Y(4660)$ via $S$-wave interaction for the discussed $e^+e^-\to D_s^{*-}D_{s2}^*(2573)^+ \to  D_s^{*-}(D^0K^+)$ process. Additionally, $D_{s2}^*(2573)^+$ dominantly decays into $D^0K^+$ through the $D$-wave interaction \cite{Kubota:1994gn,Song:2015nia,Godfrey:2015dva}. In fact, this is the main reason why we would like to firstly select this process as an example. Corresponding to Fig.  \ref{fig:1}, we identify $A\equiv Y(4660)$, $B\equiv D_s^{*-}$, $C\equiv D_{s2}^*(2573)^+$, $D\equiv D^0$, and $E\equiv K^+$.

In order to calculate this process, we adopt the effective Lagrangian approach, where the related Lagrangians include \cite{Bauer:1975bv,Bauer:1975bw,Liu:2020ruo}
\begin{eqnarray}
\mathcal{L}_{\gamma Y}&=&\frac{-em_Y^2}{f_Y}Y_{\mu}A^{\mu},\nonumber\\
\mathcal{L}_{D_{s2}{D}_s^*Y}&=&g_{D_{s2}D_s^*Y}Y_{\mu}(D_{s2}^{\mu\nu\dag}D^*_{s\nu}+D^{*\dag}_{s\nu}D_{s2}^{\mu\nu}), \nonumber \\
\mathcal{L}_{KD{D}_{s2}}&=&g_{KDD_{s2}}(D_{s2}^{\mu\nu\dag}(\partial_{\mu}\partial_{\nu}K)D+D^{\dag}(\partial_{\mu}\partial_{\nu}K)D_{s2}^{\mu\nu}).
\end{eqnarray}
Here, $A^{\mu}$, $D_{s2}^{\mu\nu}$, $K$ are photon, charmed-strange meson $D_{s2}^*(2573)$, and kaon fields, respectively, and $g_{D_{s2}D_s^*Y}$ and $g_{KDD_{s2}}$ are the coupling constants.
Thus, the general amplitude of the $e^+(k_1)e^-(k_2)\to Y(p_1) \to D_s^{*-}(p_4)D_{s2}^*(2573)^+ \to  D_s^{*-}(D^0(p_3)K^+(p_2))$ channel can be written as
\begin{eqnarray}
\mathcal{M}&=&\bar{v}(k_2)e\gamma_{\mu}u(k_1)\frac{-g^{\mu\nu}em_{Y}^2}{sf_{Y}}\frac{-g_{\nu\rho}+p_{1\nu}p_{1\rho}/m_Y^2}{s-m_Y^2+im_Y\Gamma_Y} \nonumber \\
&&\times g_{D_{s2}D_s^*Y}g_{KDD_{s2}}\epsilon_{D_s^*}^{\lambda *}\frac{-G_{\rho\lambda\alpha\beta}p_2^{\alpha}p_2^{\beta}}{(p_1-p_4)^2-m_{D_{s2}}^2+im_{D_{s2}}\Gamma_{D_{s2}}},
\end{eqnarray}
where $G_{\rho\lambda\alpha\beta}=\frac{1}{2}(\tilde{g}_{\rho\alpha}\tilde{g}_{\lambda\beta}+\tilde{g}_{\rho\beta}\tilde{g}_{\lambda\alpha})-\frac{1}{3}\tilde{g}_{\rho\lambda}\tilde{g}_{\alpha\beta}$ with $\tilde{g}_{\rho\alpha}=-g_{\rho\alpha}+(p_{1\rho}-p_{4\rho})(p_{1\alpha}-p_{4\alpha})/m_{D_{s2}}^2$ is the spin projection operator of tensor particle.
With the above preparation, the differential cross section vs. invariant mass $m_{D_s^*D}$ or the recoil spectrum of $K^+$ can be calculated by integrating over the phase space of three-body final states, i.e.,
\begin{eqnarray}
\frac{d\sigma}{dm_{D_s^*D}}=\frac{1}{(2\pi)^5}\frac{1}{32\sqrt{s}\sqrt{(k_1\cdot k_2)^2}}\overline{\left|\mathcal{M}\right|}^2\left|\bf{p_2}\right|\left|\bf {p_3}^*\right| d\Omega_{2}d\Omega_{3}^*,
\end{eqnarray}
where the overline above the amplitude stands for the average over the spin of electron and positron and the sum over the spin of final states, and the $\bf {p_3}^*$ and $\Omega_{3}^*$ are the three-momentum and solid angle of $D^0$ meson in the center of mass frame of the $D_s^*D$ system.

\begin{figure}[t]
	\includegraphics[width=8.5cm,keepaspectratio]{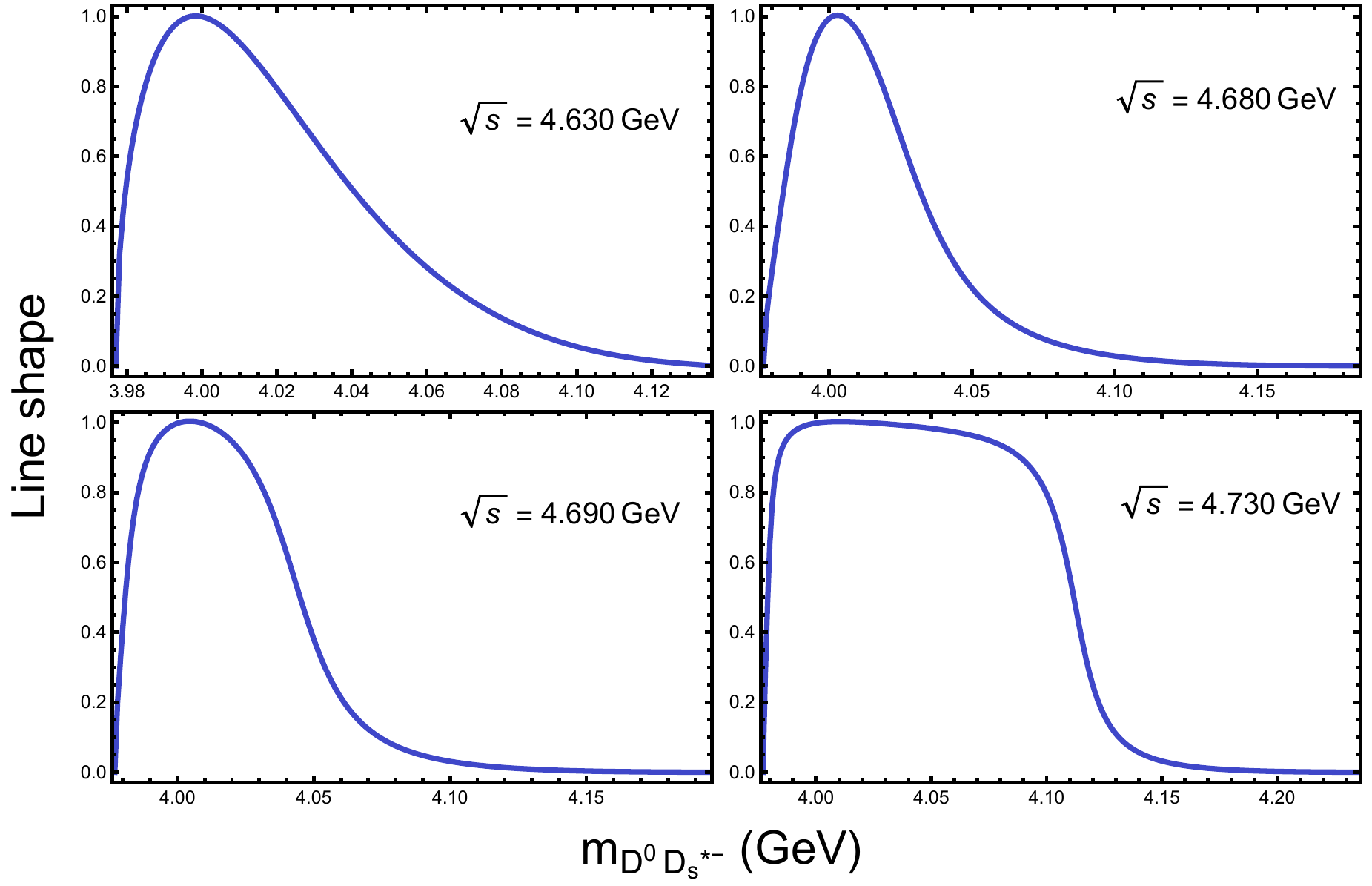}
	\caption{The line shapes of a differential cross section for $e^+e^-\to D_s^{*-}D_{s2}^*(2573)^+ \to  D_s^{*-}(D^0K^+)$ vs. the invariant mass $m_{D^0D_s^{*-}}$ with four CM energies $\sqrt{s}=4.630, 4.680, 4.690$, and $4.730$ GeV. Here, the resonance parameter of $D_{s2}^*(2573)$ is fixed as $m=2.568$ GeV and $\Gamma=16.9$ MeV \cite{Tanabashi:2018oca,Aaij:2014xza}. The maximum of line shape is normalized to 1. \label{fig:3}  }
\end{figure}

\begin{figure}[t]
	\includegraphics[width=8.0cm,keepaspectratio]{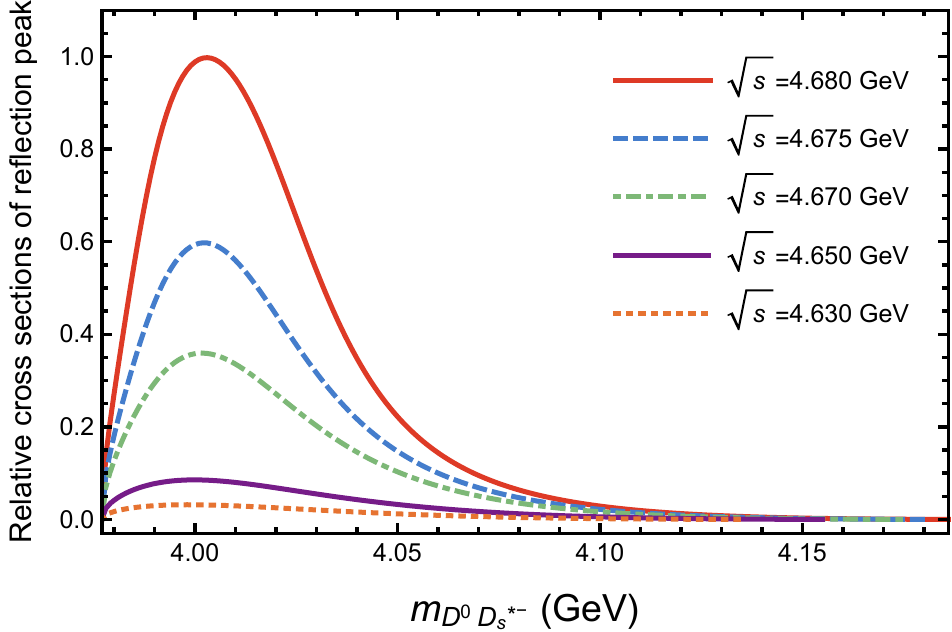}
	\caption{The relative differential cross sections of reflection peaks on the invariant mass spectrum of $D^0D_s^{*-}$ in  $e^+e^-\to D_s^{*-}D_{s2}^*(2573)^+ \to  D_s^{*-}(D^0K^+)$ with five different CM energies. The maximum of the reflection peak at $\sqrt{s}=4.680$ GeV is normalized to 1. \label{fig:5}  }
\end{figure}

In Fig. \ref{fig:3}, we test the sensitivity of the reflection line shape of $D_{s2}^*(2573)$ near the critical CM energy by plotting a differential cross section vs. the $m_{D^0D_s^{*-}}$ invariant mass
with four CM energies $\sqrt{s}=4.630, 4.680, 4.690$, and $4.730$ GeV. Here, $\sqrt{s}=4.680$ GeV is just the critical energy value for the discussed $e^+e^-\to D_s^{*-}D_{s2}^*(2573)^+ \to  D_s^{*-}(D^0K^+)$. Due to the peculiar kinematic behavior, it can be found that the line shape of a reflection signal is 
sensitive to the CM energy value around $4.680$ GeV. When $\sqrt{s}>4.680$ GeV, 
the line shape of the reflection does not look like a peak structure and becomes more flat with increasing $\sqrt{s}$. When CM energy is exactly at 4680 MeV, the line shape of reflection represents a {critical} peak structure.  {\ For another region of $\sqrt{s}< 4.680$ GeV, although all reflection line shapes look like a peak structure, their sensitivity is reflected on the relative magnitude of events because of the off-shell suppression effect of the pole. In Fig. \ref{fig:5}, the relative differential cross sections based on five CM energy points of $\sqrt{s}=4.680, 4.675, 4.670, 4.650$, and 4.630 GeV are given. It can be seen that the relative magnitude of differential cross sections is vey sensitive to the CM energy.  For example, there is about 40\% suppression for the tiny energy change of 5 MeV away from the critical energy. Additionally, there is even an order of magnitude difference between $\sqrt{s}=4.680$ and 4.650 GeV, which means that these reflection peaks that are far from the critical energy are most likely invisible in experiments. Hence, it can be seen that the reflective peak near a threshold is the most obvious at critical position of $\sqrt{s}=\sqrt{s_{\textrm{critical}}}$, which should be the best CM energy to induce the formation of the charmoniumlike structures as a reflection nature.}

\begin{figure}[t]
	\includegraphics[width=8.6cm,keepaspectratio]{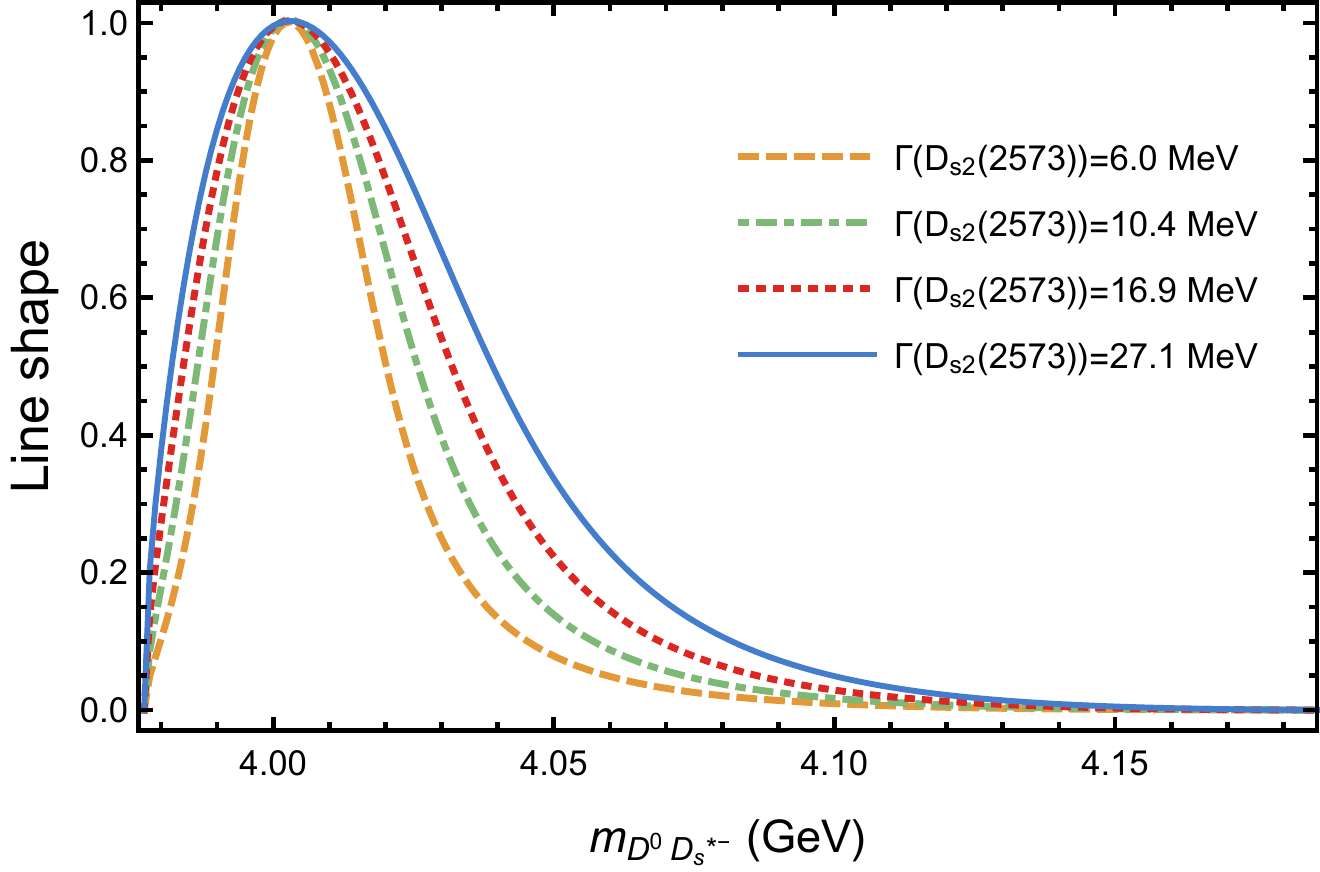}
	\caption{The dependence of the line shape of a differential cross section for $e^+e^-\to D_s^{*-}D_{s2}^*(2573)^+ \to  D_s^{*-}(D^0K^+)$ vs. the invariant mass $m_{D^0D_s^{*-}}$ 
on the width of $D_{s2}^*(2573)^+$. Here, the CM energy is $\sqrt{s}=4.680$ GeV and four typical values of width of $D_{s2}^*(2573)^+$ are 6.0, 10.4, 16.9, 27.1 MeV, which are from Ref. \cite{Song:2015nia}, ARGUS \cite{Albrecht:1995qx}, LHCb \cite{Aaij:2014xza}, and {\it BABAR} \cite{Aubert:2006mh}, respectively. Here, The maximum of line shape is normalized to 1. \label{fig:4}} %Additionally, we also compare our result with the BESIII experimental data \cite{1830518}. 
\end{figure}

Besides this, the reflection line shape is also sensitive to the width of an intermediate state $D_{s2}^*(2573)$ as shown in Fig. \ref{fig:4}, where  
all line shapes have the same pole position at $m_{D^0D_s^{*-}}^{pole}=4.003$ GeV when the CM energy is $\sqrt{s}=4.680$ GeV.
In fact, until now the resonance parameters of $D_{s2}^*(2573)$, especially its width, have not been well established by directly analyzing the corresponding Breit-Wigner distribution 
\cite{Tanabashi:2018oca}. The measured width of $D_{s2}^*(2573)$ is from 2 to 33 MeV if considering experimental uncertainties \cite{Tanabashi:2018oca}. Here, we select four typical values of width as 6.0, 10.4, 16.9, and 27.1 MeV as an example, which are from Ref. \cite{Song:2015nia}, ARGUS \cite{Albrecht:1995qx}, LHCb \cite{Aaij:2014xza}, and {\it BABAR} \cite{Aubert:2006mh}, respectively.  It can be found that the larger width of intermediate $D_{s2}^*(2573)$ is also related to a wider reflection peak structure, and its line shape is very sensitive to an intermediate resonance width, which can be seen within the width change of only several MeV.

Therefore, this critical energy induced enhancement mechanism existing in a three-body open-charm process means that experimentalists can easily implement to search for a critical reflection peak by scanning a series of the CM energy points, which is expected to be somewhere in the middle of a series of scanned CM energy points. At the same time, because of the existence of the CEE mechanism, these induced reflection structures as a new cluster of charmoniumlike structures can be more easily distinguished  from the charmoniumlike states with a resonant nature.

%The experimental study of the recoil mass spectrum of the $e^+e^-\to   D_s^{*-}(D^0K^+)$ process at $\sqrt{s}=4.680$ GeV will be a novel approach to constrain the width of $D_{s2}^*(2573)$. Thus, we suggest our BESIII colleagues to collect more data of $e^+e^-\to   D_s^{*-}(D^0K^+)$ process at $\sqrt{s}=4.680 $ GeV. 

%We notice a recent BESIII interesting result of $e^+e^-\to  D_s^{*-}(D^0K^+)$ at  the CM energy $\sqrt{s}=4.681$ GeV, where the recoil mass spectrum of $K^+$ is given \cite{1830518}. Here, we simply make a comparison of our theoretical result with this BESIII result. It is obvious that experimental data around 4 GeV can be depicted as the reflection peak resulted by $D_{s2}^*(2573)$, where the width of $D_{s2}^*(2573)$ is taken 6 GeV \cite{Song:2015nia} as input

\iffalse

{ We notice that BESIII once argued that the contribution of $D_{s2}^*(2573) \to D^0K^+$ calculated by a simple Monte Carlo simulation could not account for the front peak shown in data in Fig. \ref{fig:4}. As theorists, we only believe the experimental data provided by the experimentalists. To some extent, exploring the underlying mechanism behind the experimental data should be left to the theorists. In this work, based on the analysis from an effective Lagrangian approach, our conclusion is found to be different from that derived from a simple MC simulation \cite{1830518}.  This difference should be clarified by the joint effort from theorists and experimentalists in the future. }

\fi

\begin{table}
  	\caption{ The proposal for inducing charmoniumlike structures through our proposed CEE mechanism. The last column shows the corresponding pole position of a reflective charmoniumlike peak in the invariant mass of $D_{(s)}^{(*)}D_{(s)}^{(*)}$. \label{tab:1} }
  	\setlength{\tabcolsep}{2.0mm}{
  	\begin{tabular}{cllccccc}
			\bottomrule[1.2pt]
			%\bottomrule[1.2pt]
	   $\sqrt{s_{\textrm{critical}}}~$\footnote{ { Since the listed critical CM energies are only the reference values, experimentalists can plan to scan the CM energy points near the suggested reference value. } }  & Recommended  process   & Involved state & $ m_{D_{(s)}^{(*)}D_{(s)}^{(*)}}^{pole} $  \\
			%\hline
			\bottomrule[1.2pt]
	    4286 & $e^+e^- \to DD^*\pi$  & $D_{1}(2420)$ ($1^+$) & 3890 \\
	    4429 & $e^+e^- \to D_sD_s^*\pi$  & $D_{s1}(2460)$ ($1^+$) & 4091   \\
	    4430 & $e^+e^- \to D_s^*D_s\pi$  & $D_{s0}(2317)$ ($0^+$) & 4092 \\
	    4431 & $e^+e^- \to D^*D^*\pi$  & $D_{1}(2420)$ ($1^+$) & 4035  \\
	    4471 & $e^+e^- \to D^*D\pi$  & $D_{2}^*(2460)$ ($2^+$) & 3910 \\
	    4474 & $e^+e^- \to DD^*\pi$ & $D^*(2600)$ ($1^-$) & 3906 \\
	    4504 & $e^+e^- \to D_sD^*K$  & $D_{s1}(2536)$ ($1^+$) &  3982 \\
	    4572 & $e^+e^- \to D_s^*D_s^*\pi$  & $D_{s1}(2460)$ ($1^+$) & 4234   \\
	    4615 & $e^+e^- \to D_sD^*K$ &  $D_s(2646)$ ($0^-$)\footnote{They are theoretically predicted charmed/charmed-strange mesons that are still missing in experiments \cite{Song:2015nia,Song:2015fha}. } &  3994 \\
	    4617 & $e^+e^- \to DD^*\pi$ & $D(2750)$ ($2^-$) & 3921 \\
	    4619 & $e^+e^- \to D^*D^*\pi$ & $D^*(2600)$ ($1^-$) &  4052 \\
	    4647 & $e^+e^- \to D_s^*D^*K$  & $D_{s1}(2536)$ ($1^+$) &  4125  \\
	    4677 & $e^+e^- \to D_sDK$ & $D^*_{s1}(2700)$  ($1^-$) &  3878\\
	    4750 & $e^+e^- \to DD^*\rho$ & $D(2885)$ ($1^+$)\footnotemark[2]  & 3889   \\
	    4758 & $e^+e^- \to D_s^*D^*K$ &  $D_s(2646)$ ($0^-$)\footnotemark[2]  & 4138  \\
	    4762 & $e^+e^- \to D^*D^*\pi$ & $D(2750)$  ($2^-$) & 4068\\
	    4820 & $e^+e^- \to D_s^*DK$ & $D^*_{s1}(2700)$  ($1^-$) &  4023\\
	    4894 & $e^+e^- \to D^*D\rho$  & $D(2884)$ ($2^+$)\footnotemark[2]  & 3914  \\
	    4895 & $e^+e^- \to D^*D^*\rho$ & $D(2885)$ ($1^+$)\footnotemark[2]  & 4034   \\
	    5013 & $e^+e^- \to DD\rho$ & $D(3148)$ ($2^-$)\footnotemark[2]  & 3812 \\
	    5116 & $e^+e^- \to D_s^*D^*K^*$  & $D_{s}( 3004)$ ($2^+$)\footnotemark[2]  & 4138    \\
	    5158 & $e^+e^- \to D^*D\rho$ & $D(3148)$ ($2^-$)\footnotemark[2]  & 3960  \\
	    5229 & $e^+e^- \to D_sDK^*$ & $D_s(3260)$ ($2^-$)\footnotemark[2]  & 3924  \\
	    5372 & $e^+e^- \to D_s^*DK^*$ & $D_s(3260)$ ($2^-$)\footnotemark[2]  & 4070  \\

			\bottomrule[1.2pt]
		\end{tabular}\label{table:1}}
  \end{table}

\noindent\textit{Proposal for future experiments.}--In the following, we list a series of three-body open-charm processes from the  positron-electron annihilation, where the CM energies suitable for 
studying the charmoniumlike structures induced by the CEE mechanism are given with the information of the involved intermediate  higher charmed/charmed-strange mesons. 
In a practical measurement, due to the ensurance of the CEE mechanism, the reflection peak structure  near a threshold should be largely produced when the scanning energy point gradually close to the critical position. Thus, for the listed three-body open charm processes, experimentalists can plan a scanning interval in advance based on our suggested CM energy point in Table \ref{tab:1}. 

%However, a natural problem is that there is no straightforward way to set exact critical CM energy in experiments without any prior knowledge of the intermediate $C$ particle for the general situation. Here, we will illustrate how to utilize the CEE mechanism in practice.

%In practical measurement, when considering the case in which the scanned CM energy points are not sufficient to determine the position of a strict critical point, then, these energy points must not be too far from the critical point due to the sensitivity of cross section suppression, as long as an obvious reflection peak near threshold can be seen.

The suggested critical CM energies in Table \ref{tab:1} can be directly obtained
when we determine the involved ground state charmed or charmed-strange meson $B$ and the intermediate state $C$, which denotes a higher state of charmed/charmed-strange meson families. These energy points can be covered by the future BESIII and BelleII experiments, and even the discussing possible Super Tau-Charm Factory in China, which is planned to operate in the CM energy of 2 to 7 GeV \cite{Luo:2019xqt}.
{Here, we mention the theoretical reference masses \cite{Song:2015nia,Song:2015fha} to calculate the suggested CM energies for each of the corresponding unobserved states $C$. } Thus, we suggest that experimentalists can plan to scan the CM energy points near the suggested reference value.
For the intermediate $C$ related to three-body open-charm processes from the $e^+e^-$ annihilation, { we select some relatively narrow states for convenience of experimental measurement}. Additionally, we give { approximate} pole positions of the predicted charmoniumlike structures as a reflection peak. 

For the $e^+e^-$ collision, its CM energy can be changed according to the corresponding physical aim. Obviously, our proposal can be an interesting research topic achievable at BESIII and BelleII. 
We hope that this proposal can stimulate experimentalists' interest to apply this novel CEE mechanism to detect more charmoniumlike $Z_{c}$ or $Z_{cs}$ structures. Of course, it is worth mentioning that  in a practical experimental measurement, the observation of the critical peak structure in CEE mechanism is also dependent on the coupling strength of reaction process. Therefore, based on this starting point, further theoretical studies involving the suggested three-body open-charm processes from the $e^+e^-$ annihilation are encouraged to consider the detailed dynamics.

\noindent\textit{Summary.}--We have proposed a {\it Critical Energy induced Enhancement} (CEE) mechanism to detecting a new cluster of charmoniumlike structures, which is based on a peculiar kinematic behavior of a three-body process from the electron and positron annihilation. With $e^+e^-\to D_s^{*-}D_{s2}^*(2573)^+ \to  D_s^{*-}(D^0K^+)$ as an example, we have presented the sensitivity of the reflective line shape from $D_{s2}^*(2573)$  existing in the $m_{D_s^{*-}D^0}$ invariant mass spectrum on the CM energy of the $e^+e^-$ system. When fixing a special CM energy $\sqrt{s}=4.680$ GeV, which is just equal to $m_{D_s^{*}}+m_{D_{s2}^*(2573)}$,  an obvious reflection peak near the threshold of $D_s^{*-}D^0$ can be produced. However, when the CM energy is gradually larger than 4.680 GeV, the reflection peak will disappear. For another case of $\sqrt{s}<4.680$ GeV, the events of reflection peak will be obviously suppressed due to the missing of the single pole. Therefore, the energy of $\sqrt{s}=m_{D_s^{*}}+m_{D_{s2}^*(2573)}$ is a special critical point and there exist a CEE mechanism for the three-body open-charm process from the $e^+e^-$ collisions.  A direct application of this mechanism  found in this work is to study the new cluster of charmoniumlike $Z_c$ or $Z_{cs}$ structures. We have further suggested a series of three-body open-charm processes from the electron and positron annihilation associated with the corresponding { suggested} CM energies. We have also noticed that BESIII released “{\it White paper on the Future Physics Program of BESIII}” recently \cite{Ablikim:2019hff}. Since the suggested CM energy points involved in three-body open-charm processes can be covered by BEPCII, it is obvious that it is a good chance for BESIII. The running BelleII is also an experimental facility based on the electron and positron collision. By the initial state radiation technique, BelleII has enough potential to measure the suggested three-body open-charm processes.  Experimentally studying three-body open-charm processes from electron and positron annihilation with special CM energy points may form a future research topic accessible at BESIII, BelleII and even Super Tau-Charm factory \cite{Luo:2019xqt}. 

In fact, this method developed in the present work is not limited only to the suggested three-body open-charm processes mentioned above. We may continue to extend it to three-body open-strange processes and three-body open-bottom processes from the electron and positron annihilation, which can be an effective approach to construct new strangeoniumlike and bottomoniumlike structures families, respectively.

\section*{ACKNOWLEDGEMENTS}

This work is partly supported by the China National Funds for Distinguished Young Scientists under Grant No. 11825503, National Key Research and Development Program of China under Contract No. 2020YFA0406400, the 111 Project under Grant No. B20063, the National Natural Science Foundation of China under Grant Nos. 12047501 and 11775050, and the Fundamental Research Funds for the Central Universities under Grant No. lzujbky-2020-it03.


\begin{thebibliography}{199}


%\cite{Chen:2016qju}
\bibitem{Chen:2016qju}
  H.~X.~Chen, W.~Chen, X.~Liu and S.~L.~Zhu,
  The hidden-charm pentaquark and tetraquark states,
  Phys.\ Rept.\  {\bf 639} (2016) 1.
%  doi:10.1016/j.physrep.2016.05.004
%  [arXiv:1601.02092 [hep-ph]].
  %%CITATION = doi:10.1016/j.physrep.2016.05.004;%%
  %499 citations counted in INSPIRE as of 08 Feb 2020

%\cite{Liu:2019zoy}
\bibitem{Liu:2019zoy}
  Y.~R.~Liu, H.~X.~Chen, W.~Chen, X.~Liu and S.~L.~Zhu,
  Pentaquark and Tetraquark states,
  Prog.\ Part.\ Nucl.\ Phys.\  {\bf 107}, 237 (2019).
%  doi:10.1016/j.ppnp.2019.04.003
%  [arXiv:1903.11976 [hep-ph]].
  %%CITATION = doi:10.1016/j.ppnp.2019.04.003;%%
  %77 citations counted in INSPIRE as of 08 Feb 2020
  
  
 %\cite{Guo:2017jvc}
\bibitem{Guo:2017jvc}
F.~K.~Guo, C.~Hanhart, U.~G.~Mei\ss{}ner, Q.~Wang, Q.~Zhao and B.~S.~Zou,
Hadronic molecules,
Rev. Mod. Phys. \textbf{90}, no.1, 015004 (2018).
%doi:10.1103/RevModPhys.90.015004
%[arXiv:1705.00141 [hep-ph]].
%530 citations counted in INSPIRE as of 16 Mar 2021


%\cite{Olsen:2017bmm}
\bibitem{Olsen:2017bmm}
S.~L.~Olsen, T.~Skwarnicki and D.~Zieminska,
Nonstandard heavy mesons and baryons: Experimental evidence,
Rev. Mod. Phys. \textbf{90}, no.1, 015003 (2018).
%doi:10.1103/RevModPhys.90.015003
%[arXiv:1708.04012 [hep-ph]].
%345 citations counted in INSPIRE as of 16 Mar 2021


%\cite{Brambilla:2019esw}
\bibitem{Brambilla:2019esw}
N.~Brambilla, S.~Eidelman, C.~Hanhart, A.~Nefediev, C.~P.~Shen, C.~E.~Thomas, A.~Vairo and C.~Z.~Yuan,
The $XYZ$ states: experimental and theoretical status and perspectives,
Phys. Rept. \textbf{873}, 1-154 (2020).
%doi:10.1016/j.physrep.2020.05.001
%[arXiv:1907.07583 [hep-ex]].
%183 citations counted in INSPIRE as of 16 Mar 2021



%\cite{Aubert:2005rm}
\bibitem{Aubert:2005rm}
B.~Aubert \textit{et al.} [BaBar],
Observation of a broad structure in the $\pi^+ \pi^- J/\psi$ mass spectrum around 4.26 GeV/c$^2$,
Phys. Rev. Lett. \textbf{95}, 142001 (2005).
%doi:10.1103/PhysRevLett.95.142001
%[arXiv:hep-ex/0506081 [hep-ex]].
%897 citations counted in INSPIRE as of 16 Mar 2021


%\cite{Yuan:2007sj}
\bibitem{Yuan:2007sj}
C.~Z.~Yuan \textit{et al.} [Belle],
Measurement of $e^+ e^- \to \pi^+ \pi^- J/\psi$ cross section via initial state radiation at Belle,
Phys. Rev. Lett. \textbf{99}, 182004 (2007).
%doi:10.1103/PhysRevLett.99.182004
%[arXiv:0707.2541 [hep-ex]].
%469 citations counted in INSPIRE as of 16 Mar 2021


%\cite{Wang:2007ea}
\bibitem{Wang:2007ea}
X.~L.~Wang \textit{et al.} [Belle],
Observation of Two Resonant Structures in $e^+e^-\to \pi^+ \pi^- \psi(2S)$ via Initial State Radiation at Belle,
Phys. Rev. Lett. \textbf{99}, 142002 (2007).
%doi:10.1103/PhysRevLett.99.142002
%[arXiv:0707.3699 [hep-ex]].
%451 citations counted in INSPIRE as of 16 Mar 2021


%\cite{Aubert:2006ge}
\bibitem{Aubert:2006ge}
B.~Aubert \textit{et al.} [BaBar],
Evidence of a broad structure at an invariant mass of 4.32  GeV/c$^{2}$ in the reaction $e^{+} e^{-} \to \pi^{+} \pi^{-} \psi(2S)$ measured at BaBar,
Phys. Rev. Lett. \textbf{98}, 212001 (2007).
%doi:10.1103/PhysRevLett.98.212001
%[arXiv:hep-ex/0610057 [hep-ex]].
%361 citations counted in INSPIRE as of 16 Mar 2021


%\cite{Pakhlova:2008vn}
\bibitem{Pakhlova:2008vn}
G.~Pakhlova \textit{et al.} [Belle],
Observation of a near-threshold enhancement in the $e^+e^- \to \Lambda^+_{c} \Lambda^-_{c} $ cross section using initial-state radiation,
Phys. Rev. Lett. \textbf{101}, 172001 (2008).
%doi:10.1103/PhysRevLett.101.172001
%[arXiv:0807.4458 [hep-ex]].
%242 citations counted in INSPIRE as of 16 Mar 2021


%\cite{BESIII:2016adj}
\bibitem{BESIII:2016adj}
M.~Ablikim \textit{et al.} [BESIII],
Evidence of Two Resonant Structures in $e^+ e^- \to \pi^+ \pi^- h_c$,
Phys. Rev. Lett. \textbf{118}, no.9, 092002 (2017).
%doi:10.1103/PhysRevLett.118.092002
%[arXiv:1610.07044 [hep-ex]].
%126 citations counted in INSPIRE as of 16 Mar 2021


%\cite{Ablikim:2016qzw}
\bibitem{Ablikim:2016qzw}
M.~Ablikim \textit{et al.} [BESIII],
Precise measurement of the $e^+e^-\to \pi^+\pi^-J/\psi$ cross section at center-of-mass energies from 3.77 to 4.60 GeV,
Phys. Rev. Lett. \textbf{118}, no.9, 092001 (2017).
%doi:10.1103/PhysRevLett.118.092001
%[arXiv:1611.01317 [hep-ex]].
%163 citations counted in INSPIRE as of 16 Mar 2021


%\cite{Ablikim:2019hff}
\bibitem{Ablikim:2019hff}
M.~Ablikim \textit{et al.} [BESIII],
Future Physics Programme of BESIII,
Chin. Phys. C \textbf{44}, no.4, 040001 (2020).
%doi:10.1088/1674-1137/44/4/040001
%[arXiv:1912.05983 [hep-ex]].
%20 citations counted in INSPIRE as of 27 May 2020


%\cite{Abe:2007jna}
\bibitem{Abe:2007jna}
K.~Abe \textit{et al.} [Belle],
Observation of a new charmonium state in double charmonium production in $e^+ e^-$ annihilation at $\sqrt{s} \approx 10.6$  GeV,
Phys. Rev. Lett. \textbf{98}, 082001 (2007).
%doi:10.1103/PhysRevLett.98.082001
%[arXiv:hep-ex/0507019 [hep-ex]].
%355 citations counted in INSPIRE as of 16 Mar 2021



%\cite{Abe:2007sya}
\bibitem{Abe:2007sya}
P.~Pakhlov \textit{et al.} [Belle],
Production of New Charmoniumlike States in $e^+ e^- \to J/\psi D^{(*)} \bar{D}^{(*)}$ at $\sqrt{s} \approx 10.6$ GeV,
Phys. Rev. Lett. \textbf{100}, 202001 (2008).
%doi:10.1103/PhysRevLett.100.202001
%[arXiv:0708.3812 [hep-ex]].
%237 citations counted in INSPIRE as of 16 Mar 2021


%\cite{Ablikim:2013xfr}
\bibitem{Ablikim:2013xfr}
M.~Ablikim \textit{et al.} [BESIII],
Observation of a charged $(D\bar{D}^{*})^\pm$ mass peak in $e^{+}e^{-} \to \pi D\bar{D}^{*}$ at $\sqrt{s} =$ 4.26 GeV,
Phys. Rev. Lett. \textbf{112}, no.2, 022001 (2014).
%doi:10.1103/PhysRevLett.112.022001
%[arXiv:1310.1163 [hep-ex]].
%312 citations counted in INSPIRE as of 16 Mar 2021


%\cite{Ablikim:2013emm}
\bibitem{Ablikim:2013emm}
M.~Ablikim \textit{et al.} [BESIII],
Observation of a charged charmoniumlike structure in $e^+e^- \to (D^{*} \bar{D}^{*})^{\pm} \pi^\mp$ at $\sqrt{s}=4.26$ GeV,
Phys. Rev. Lett. \textbf{112}, no.13, 132001 (2014).
%doi:10.1103/PhysRevLett.112.132001
%[arXiv:1308.2760 [hep-ex]].
%333 citations counted in INSPIRE as of 16 Mar 2021


%\cite{Wang:2020axi}
\bibitem{Wang:2020axi}
J.~Z.~Wang, D.~Y.~Chen, X.~Liu and T.~Matsuki,
Universal non-resonant explanation to charmoniumlike structures $Z_c(3885)$ and $Z_c(4025)$,
Eur. Phys. J. C \textbf{80}, no.11, 1040 (2020).
%doi:10.1140/epjc/s10052-020-08621-4
%[arXiv:2007.02263 [hep-ph]].
%2 citations counted in INSPIRE as of 16 Mar 2021





%\cite{Tanabashi:2018oca}
\bibitem{Tanabashi:2018oca}
M.~Tanabashi \textit{et al.} [Particle Data Group],
Review of Particle Physics,
Phys. Rev. D \textbf{98}, no.3, 030001 (2018).
%doi:10.1103/PhysRevD.98.030001
%4697 citations counted in INSPIRE as of 27 May 2020



%\cite{Aaij:2014xza}
\bibitem{Aaij:2014xza}
R.~Aaij \textit{et al.} [LHCb],
Observation of overlapping spin-1 and spin-3 $\bar{D}^0 K^-$ resonances at mass $2.86 {\rm GeV}/c^2$,
Phys. Rev. Lett. \textbf{113}, 162001 (2014).
%doi:10.1103/PhysRevLett.113.162001
%[arXiv:1407.7574 [hep-ex]].
%90 citations counted in INSPIRE as of 27 May 2020



%\cite{Wang:2020prx}
\bibitem{Wang:2020prx}
J.~Z.~Wang, R.~Q.~Qian, X.~Liu and T.~Matsuki,
Are the $Y$ states around 4.6 GeV from $e^+e^-$ annihilation higher charmonia?,
Phys. Rev. D \textbf{101}, no.3, 034001 (2020).
%doi:10.1103/PhysRevD.101.034001
%[arXiv:2001.00175 [hep-ph]].
%3 citations counted in INSPIRE as of 27 May 2020

%\cite{Kubota:1994gn}
\bibitem{Kubota:1994gn}
Y.~Kubota \textit{et al.} [CLEO],
Observation of a new charmed strange meson,
Phys. Rev. Lett. \textbf{72}, 1972-1976 (1994).
%doi:10.1103/PhysRevLett.72.1972
%[arXiv:hep-ph/9403325 [hep-ph]].
%84 citations counted in INSPIRE as of 27 May 2020

%\cite{Song:2015nia}
\bibitem{Song:2015nia}
Q.~T.~Song, D.~Y.~Chen, X.~Liu and T.~Matsuki,
Charmed-strange mesons revisited: mass spectra and strong decays,
Phys. Rev. D \textbf{91}, 054031 (2015).
%doi:10.1103/PhysRevD.91.054031
%[arXiv:1501.03575 [hep-ph]].
%38 citations counted in INSPIRE as of 27 May 2020

%\cite{Godfrey:2015dva}
\bibitem{Godfrey:2015dva}
S.~Godfrey and K.~Moats,
Properties of Excited Charm and Charm-Strange Mesons,
Phys. Rev. D \textbf{93}, no.3, 034035 (2016).
%doi:10.1103/PhysRevD.93.034035
%[arXiv:1510.08305 [hep-ph]].
%81 citations counted in INSPIRE as of 27 May 2020

%\cite{Bauer:1975bv}
\bibitem{Bauer:1975bv}
T.~Bauer and D.~Yennie,
Corrections to VDM in the Photoproduction of Vector Mesons. 1. Mass Dependence of Amplitudes,
Phys. Lett. B \textbf{60}, 165-168 (1976).
%doi:10.1016/0370-2693(76)90414-7
%16 citations counted in INSPIRE as of 27 May 2020

%\cite{Bauer:1975bw}
\bibitem{Bauer:1975bw}
T.~Bauer and D.~Yennie,
Corrections to Diagonal VDM in the Photoproduction of Vector Mesons. 2. Phi-omega Mixing,
Phys. Lett. B \textbf{60}, 169-171 (1976).
%doi:10.1016/0370-2693(76)90415-9
%35 citations counted in INSPIRE as of 27 May 2020

%\cite{Liu:2020ruo}
\bibitem{Liu:2020ruo}
J.~Liu, Q.~Wu, J.~He, D.~Y.~Chen and T.~Matsuki,
Production of $P-$wave charmed and charmed-strange mesons in pion and kaon induced reactions,
Phys. Rev. D \textbf{101}, no.1, 014003 (2020).
%doi:10.1103/PhysRevD.101.014003
%[arXiv:2001.00212 [hep-ph]].
%0 citations counted in INSPIRE as of 27 May 2020

%\cite{Albrecht:1995qx}
\bibitem{Albrecht:1995qx}
H.~Albrecht \textit{et al.} [ARGUS],
Measurement of the decay $D_{s2}^{*+} \to  D^0 K^+$,
Z. Phys. C \textbf{69}, 405-408 (1996).
%doi:10.1007/s002880050040
%18 citations counted in INSPIRE as of 27 May 2020

%\cite{Aubert:2006mh}
\bibitem{Aubert:2006mh}
B.~Aubert \textit{et al.} [BaBar],
Observation of a New $D_s$ Meson Decaying to $DK$ at a Mass of 2.86 GeV/$c^2$,
Phys. Rev. Lett. \textbf{97}, 222001 (2006).
%doi:10.1103/PhysRevLett.97.222001
%[arXiv:hep-ex/0607082 [hep-ex]].
%196 citations counted in INSPIRE as of 27 May 2020



%\cite{Song:2015fha}
\bibitem{Song:2015fha}
Q.~T.~Song, D.~Y.~Chen, X.~Liu and T.~Matsuki,
Higher radial and orbital excitations in the charmed meson family,
Phys. Rev. D \textbf{92}, no.7, 074011 (2015).
%doi:10.1103/PhysRevD.92.074011
%[arXiv:1503.05728 [hep-ph]].
%31 citations counted in INSPIRE as of 27 May 2020



%\cite{Luo:2019xqt}
\bibitem{Luo:2019xqt}
Q.~Luo, W.~Gao, J.~Lan, W.~Li and D.~Xu,
Progress of Conceptual Study for the Accelerators of a 2-7GeV Super Tau Charm Facility at China,
 in Proceedings of 10th International Particle Accelerator Conf. (IPAC’19), Melbourne, Australia, 2019.
%(JACoW Publishing, Geneva, 2019).
%doi:10.18429/JACoW-IPAC2019-MOPRB031
%2 citations counted in INSPIRE as of 27 May 2020






\end{thebibliography}
\end{document}